\documentstyle[multicol,aps]{revtex}
\begin{document}
\draft

\title{Charge fluctuations and the tunneling spectra of
non-magnetic metallic nanoparticles }

\author{Gustavo A. Narvaez and George Kirczenow}

\address{Department of Physics, Simon Fraser University, Burnaby,
British Columbia, Canada, V5A 1S6}

\date{\today}
\maketitle

\begin{abstract}

We present microscopic transport calculations of the tunneling spectra of
non-magnetic metal nanoparticles. We show that charge fluctuations give rise to
tunneling resonances of a new type. Positive and  negative
fluctuations have differing kinetics and thus account for previously
unexplained spectral features that are found experimentally under only
forward or only reverse applied bias. The
observed  clustering of tunneling resonances of Al nanoparticles arises
naturally from our theory.

\end{abstract}

\begin{multicols}{2}

In metal particles with dimensions on the nanometer
scale the electron de Broglie wavelength is
comparable to the size of the particle. Thus these nanoparticles
exhibit discrete electronic spectra that can be observed directly
in tunneling measurements.\cite{reviews_topic} Such
experiments have recently been carried out for
Al,\cite{ralph_experiments} Au, Ag, Cu
\cite{nonmag_experiments} and Co \cite{cobalt_experiments} nanoparticles
coated with aluminum oxide that forms the tunnel barrier. They have attracted
considerable attention since they can in principle provide detailed
microscopic information relevant to many important but poorly understood
aspects of nanoscale metal physics that range from the effects of disorder and
surface chemistry to nanoscale ferromagnetism and superconductivity.
\cite{reviews_topic}
Some of the observed effects have been modelled phenomenologically with
considerable success.
\cite{Agam_PRL_1997,canali_PRL_2000,kleff_PRB_2001,aleiner_PhysRep_2002,Bonet_PRB_2002}
However even in the case of Al nanoparticles (the simplest and most studied of
these systems) the present understanding of the results of the experiments is
far from satisfactory. For example, some of the tunneling resonances that are
seen experimentally can be matched with similar features that are observed
when the bias applied to the nanoparticle is reversed.
Thus they can reasonably be attributed to tunneling
through particular electronic states of
the nanoparticle.\cite{ralph_experiments} However,
other observed tunneling resonances have no
identifiable counterparts under reverse bias and their physical origin has
remained a mystery.\cite{ralph_experiments} This suggests that some
important aspects of the
physics of electron transport through the metal nanoparticles have
not been recognized
to date. In this Communication we identify a plausible candidate: We
demonstrate theoretically that {\em charge fluctuations} that occur
whenever a current flows through the nanoparticle should result in
transport resonances of a new type that begin already {\em in the first
step} of the Coulomb staircase. We predict that for most samples these
new resonances (unlike other tunneling features) should be
{\em much} stronger for one direction of the applied bias
than for the other. Thus charge fluctuations account
naturally for the presence in the experimental tunneling
spectra of the previously unexplained resonances described
above. We argue that in typical samples this new
mechanism should account for a substantial fraction of all
of the observed tunneling features.

We illustrate our
predictions with numerical calculations for Al nanoparticles
whose electronic structure is described by a
microscopic tight-binding model
\cite{NarvaezKirczenow_PRB_2002} that incorporates the geometry of the
particle and accounts for the presence of disorder as well as the detailed
chemistry of the  metal-oxide interface. Salient features of the electronic
structure are depicted in Fig. \ref{Fig_1} which shows the
calculated energy eigenvalues
$E_i$ near the Fermi level $E_F$ of a disc-shaped
nanoparticle of volume
${\cal V}= 16.9 nm^3$. The
amplitudes of the electron eigenfunctions
$\varphi_i$ on the top ($T$) and bottom ($B$) surfaces of the nanoparticle
          are also displayed for selected electron eigenstates.
The diameter of each circle represents the magnitude of
$\varphi_i$ at a given atomic site
(indicated by the central dot). Due to the presence of surface
disorder, the relief of the
amplitude is quite complex and depends strongly on the electronic
state that is considered and
on which surface is shown.

%
%
%
%

The wavefunction landscapes at these surfaces enter the
electron tunneling efficiencies $\gamma^{\lambda}_i$ between level
$i$ of the nanoparticle and contact $\lambda$ (=$T$,$B$) as follows:
Let $M^{\lambda}_{e,i}$ be
the tunneling matrix element  between the
electronic state
$\Psi^{\lambda}_e(\vec{r})=\langle \vec{r}|\Psi^{\lambda}_e\rangle$
of the contact and the
state $\varphi_i(\vec{r})=\langle\vec{r}|\varphi_i\rangle$ of the
nanoparticle. Since $M^{\lambda}_{e,i}$ is a transfer matrix element
we adopt the expression
$M^{\lambda}_{e,i}= {\cal R}\int_{\Omega_{\lambda}} d\vec{S}\,
\Psi^{\lambda}_e
\varphi_i$
widely used in the quantum chemistry literature to calculate
transfer matrix elements between different molecular
states.\cite{mcglynn_book}
$\cal{R}$ is an energy scale factor and
$\Omega_{\lambda}$ is the $T$ or $B$ surface of the  nanoparticle.
The nanoparticle electron wavefunction is
$\varphi_i(\vec{r})=\sum_{\vec{R}_j}a^{\alpha}_{j,i}\phi_{\alpha}(\vec{r}-\vec{R}_j)$
where the coefficients $a^{\alpha}_{j,i}$ arise from diagonalization of a
tight-binding Hamiltonian constructed
using a {\em s}, {\em p},
{\em d} orbital basis ($\phi_\alpha$).\cite{NarvaezKirczenow_PRB_2002}
Obtaining a realistic $\Psi^{\lambda}_e(\vec{r})$ in the contact/oxide
region is difficult. However, we do not need the whole wavefunction but
only the values that $\Psi^{\lambda}_e(\vec{r})$ takes at the surface of
the nanoparticle.
         From a tight-binding point of view with only {\em s} orbitals in
the basis,
the wavefunction at the surface can be written
$\Psi^{\Omega_{\lambda}}_e(\vec{r})=\sum_{\vec{r}_j}b^{s}_{j,e}
\phi_s (\vec{r}-\vec{r}_j)$ with $\vec{r}_j$ the positions of the atomic sites
at $\Omega_{\lambda}$.  Due to the disordered nature of the
metal-oxide interface and time reversal symmetry, we
represent the coefficients $b^{s}_{j,e}$ by those arising from
diagonalization of a matrix
${\cal M}$ of order ${\cal N}$ within the Gaussian Orthogonal Ensemble
weighted by the WKB extinction coefficient $e^{-\kappa
d^{ox}_{\lambda}}.$\cite{schiff_book_1955}
${\cal N}$ is the number of sites at surface
$\Omega_{\lambda}$,  $\kappa =\sqrt{2mV_b/\hbar^2}$ and
$V_b \simeq 1.2 eV$ \cite{Rippard_PRL_2002};
$d^{ox}_{\lambda}$ is the thickness of the oxide barrier.
We average over an ensemble of matrices
${\cal M}$ to model the fact that different lead states
$|\Psi_e\rangle$ couple to a given nanoparticle state $|\varphi_i\rangle$.
Thus we find

\begin{equation}
\label{Eq_gammas}
\gamma^{\lambda}_i=\frac{2\pi}{\hbar} {\cal R}^2 \nu_{\lambda}(E_H)
e^{-2\kappa
d^{ox}_{\lambda}}\Big\langle\Big|\sum_{\vec{R}_j}\sum_{\alpha}[b^{s}_{j,e}
a^{\alpha}_{j,i}]_{_{\lambda}}\Big|^2\Big\rangle.
\end{equation}

\noindent $ \nu_{\lambda}(E_H)$ is the density of states at the highest
occupied level in the lead $\lambda$ and $\langle\cdots \rangle$ denotes
ensemble averaging.

In Eq. \ref{Eq_gammas} $\gamma^{\lambda}_i$ depends on the oxide
thickness $d^{ox}_{\lambda}$ between the $\lambda$-lead and the
nanoparticle. Within a simple parallel-plate capacitor model this
thickness also determines the lead-dot capacitances ${\cal C}_{T}$ and ${\cal
C}_B$ for leads $T$ and $B$; for simplicity we assume
${\cal C}_{T}d^{ox}_{T}={\cal C}_{B}d^{ox}_{B}$. Fig.
\ref{Fig_2} shows the calculated $\gamma^{\lambda}_i$  , in
units of $\Gamma_0=2\pi {\cal R}^2 \nu_{\lambda}(E_H)/ \hbar$ for the
$T$ and $B$
surfaces, for the first few energy levels around the Fermi level for
different capacitance ratios (${\cal C}_{B}/{\cal C}_{T}=1,1.25$). We take
as a reference $d^{ox}_{B}=5\AA$ and $\kappa=0.56\AA^{-1}$\cite{note_0}.
When the thicknesses (capacitances) are equal, the tunneling
efficiencies, although asymmetric, are of the same order of magnitude.
As soon as the thickness (capacitance) ratio is changed {\em slightly}
so that
$d^{ox}_{B} < d^{ox}_{T}$, the asymmetry becomes stronger and
$\gamma^T_i$ falls rapidly by almost an order of magnitude relative to
$\gamma^B_i$.


We calculate the electric current $I$ through the nanoparticle and the
differential conductance $dI/dV$ assuming for simplicity that when an
electron enters or leaves the nanoparticle the latter relaxes quickly
to its electronic ground state. The complementary limit of slow
relaxation has been studied previously
\cite{Agam_PRL_1997,Bonet_PRB_2002}, omitting however the
new effects of charge fluctuations that we introduce here.
We consider forward bias
(FB) and reverse  bias (RB) voltages  $V$ at which the
electron population
$N$ of the nanoparticle changes by no more than
$\pm 1$ from its neutral value $N_0$.
           Thus
the nanoparticle is neutral ($ n = N-N_0 = 0$) or negatively ($  n
=  1$) or positively ($  n =  -1$) charged. We treat the electrostatic
charging energy of the nanoparticle and leads according to standard
Coulomb blockade theory\cite{russians_metal_grains}.
            We assume that at $V=0$  the Fermi levels of the
leads align with the highest occupied level
of the neutral particle. The electrochemical
potential of the $T(B)$ lead is
$\mu^{T(B)}=E_{F} \pm e V {\cal C}_{B(T)}/{\cal{C}}_{\Sigma}$ with
${\cal{C}}_{\Sigma}={\cal{C}}_{T}+{\cal{C}}_{B}$\cite{note_1}.
The master equation for the population of the
nanoparticle is then

\end{multicols}


\begin{equation}
\label{Eq_1_master}
{\begin{array}{ll}
\partial_t   n = &
\delta_{n,0}\Big\{\sum_{i}\gamma^{T}_{i}f(E_i+U-\mu^T)
\sum_{\sigma} [1-\theta(E_i-E^{\sigma}_F)]
-\sum_{j}\gamma^{B}_{j}[1-f(E_j-U-\mu^B)]\sum_{\sigma}
\theta(E_i -E^{\sigma}_F)\Big\}
\\
\\
           &
+\delta_{n,-1}\sum_{k}[\gamma^{T}_{k}f(E_k-U-\mu^T)+
\gamma^{B}_{k}f(E_k-U-\mu^B)]\sum_{\sigma,\sigma^{\prime}}
[1-\theta(E_k-E_F^{\sigma,-\sigma^{\prime}})]
\\ \\
           &  -\delta_{n,+1}\sum_{l}\{\gamma^{B}_{l}[1-f(E_l+U-\mu^B)]+
\gamma^{T}_{l}[1-f(E_l+U-\mu^T)]\}
\sum_{\sigma,\sigma^{\prime}}\theta(E_l-E_F^{\sigma,+\sigma^{\prime}})
\\
\\ & = \delta_{n,0}({\cal S}^{T\rightarrow d}_1-{\cal S}^{d\rightarrow B}_2) +
\delta_{n,-1} ({\cal S}^{T\rightarrow d}_3+{\cal S}^{B\rightarrow d}_3)-
\delta_{n,+1}({\cal S}^{d\rightarrow B}_4+{\cal S}^{d\rightarrow T}_4),
\end{array}}
\end{equation}

\begin{multicols}{2}

\noindent where the $E_i$ are the one-electron energy levels of the neutral
nanoparticle, $U=e^2/2{\cal C}_{\Sigma}$ is the single-electron charging
energy \cite{russians_metal_grains},
$f(x)$ is the Fermi distribution of the leads at temperature
$T_d$ and $\theta(x)$ is the Heaviside function describing the occupation of
the levels in the nanoparticle. $E^{\sigma}_F$
is the highest occupied level with spin $\sigma(=\uparrow,\downarrow$).
$E_F^{\sigma,-\sigma^{\prime}}$ and
$E_F^{\sigma,+\sigma^{\prime}}$ denote the highest occupied level
of the nanoparticle with spin $\sigma$ after removal($-$) or
addition(+) of an electron with spin $\sigma^{\prime}$.
In Eq.
\ref{Eq_1_master} we assume FB with electron flow
from $T$ to $B$, therefore  $\mu^T>\mu^B$. In
RB the indices $T$ and
$B$ are interchanged.

We now discuss the physical meaning of the terms in Eq.
\ref{Eq_1_master}, keeping for the sake of clarity to
the case
$T_d=0$. The sum over
$i$ in Eq.
\ref{Eq_1_master} describes processes in which an electron moves from lead
$T$ to the single-electron state $|\varphi_i\rangle$ of an initially
neutral ($n=0$) nanoparticle. This is
allowed energetically if
$E_i+U \leq E_F+e{\cal C}_B V/{\cal C}_{\Sigma}$,
i.e., the single-electron energy $E_i$ of the level in the dot to which the
transition is made plus the charging energy
$U$ must be lower than  $\mu^T$, the highest occupied level in the
electrode. Also
the single-particle state
$|\varphi_i\rangle$ of the nanoparticle that accepts the electron, must be
initially unoccupied. $f(x)$ and  $\theta (x)$ with
the arguments in Eq.
\ref{Eq_1_master} account for these constraints.
Analogous reasoning leads  to the other
terms in  Eq. \ref{Eq_1_master}.
The sum over $j$ describes tunneling from the neutral particle to
electrode $B$ which is allowed energetically if
%
$E_j-U \geq \mu^B$.
%
If the nanoparticle is {\em initially  charged} different
energetic restrictions apply:
If $n=-1$ initially, tunneling from electrode $T(B)$ to the nanoparticle
state $|\varphi_k\rangle$ is allowed if
$E_{k}-U  \leq \mu^{T(B)}$.
If
$n=1$ initially, an electron may tunnel from
state $|\varphi_l\rangle$ of the nanoparticle
to electrode
$B(T)$ if
%
$E_l+U \geq \mu^{B(T)}$.
%
%

           The rate equations for the probability of the nanoparticle
being in a given charge state ($n=-1,0,+1$)  are:
$\partial_t P_1={\cal S}_1 P_0 - {\cal S}_4 P_{1}$, and
$\partial_t P_{-1} = {\cal
S}_2 P_0 - {\cal S}_3 P_{-1}$,
with $ P_0 +  P_{1} +  P_{-1} =1 $. In {\em forward} bias the
total rates are
${\cal S}_1= {\cal S}^{T\rightarrow d}_1$,
${\cal S}_2= {\cal S}^{d\rightarrow B}_2$, ${\cal
S}_3 = {\cal S}^{T\rightarrow d}_3+{\cal S}^{B\rightarrow d}_3 $, and ${\cal
S}_4 = {\cal S}^{d\rightarrow B}_4+{\cal S}^{d\rightarrow T}_4$.   We
solve these rate equations for the steady state occupation probabilities
$P^{st}_n$. Then the current
${\cal I}$ passing through the nanoparticle is

\begin{equation}
{\cal I}^{FB}=e ( {\cal S}^{T\rightarrow d}_1 P^{st}_0 + {\cal
S}^{T\rightarrow d}_3 P^{st}_{-1} - {\cal S}^{d\rightarrow T}_4 P^{st}_{1}
           ).
\label{Eq_current}
\end{equation}

Figure \ref{Fig_3} shows our calculated differential conductance
spectra for Al nanoparticles within the first step of the
Coulomb staircase. The
resonances labelled $i \lambda$ ($\lambda i$)  are due to tunneling
from  (to) level $|E_F+i\rangle$ of the {\em neutral}
nanoparticle to (from) contact $\lambda=T,B$. Here $\bar{i} = -i$.
We refer to these resonances as {\em neutral peaks} (NP).
        The other peaks labelled $Q^{+} (\bar{Q})$
are due to the $n= 1$ ($-1$) charge fluctuations on the nanoparticle
that are induced  by the current.
They occur when a level of the charged nanoparticle
(renormalized by the charging energy $U$)
becomes available for transport
as the electrochemical potentials of the contacts are swept.
These resonances will be referred to
as {\em charge fluctuation peaks} (CFP).
Since on the first step
of the Coulomb staircase the nanoparticle is neutral most of the time,
most NP's are much stronger than most CFP's.
However as will be explained below,
some CFP's can be strongly enhanced by kinetic bottlenecks and thus become
very prominent spectral features.

Another striking aspect of Fig. \ref{Fig_3}
is the strong tendency of the resonances  to cluster that is also found
experimentally\cite{ralph_experiments}. To account for
this clustering, previous theories \cite{Agam_PRL_1997,Bonet_PRB_2002}
employed the {\em phenomenological} assumption that
excited nanoparticle states are very long lived.
Here we do {\em not} make this assumption; the
clustering in Fig. \ref{Fig_3} follows directly from our {\em microscopic}
theory\cite{NarvaezKirczenow_PRB_2002} of the electronic structure of
the nanoparticle. Also
in common with the experimental data\cite{ralph_experiments}
(but unlike the results of previous
theories\cite{Agam_PRL_1997}) the amplitudes of the
peaks in Fig. \ref{Fig_3}
show no systematic decrease with increasing bias $V$.

We now discuss the tunneling spectra in Fig. \ref{Fig_3} in more detail.
For FB we assume $\mu^{T} > \mu^{B}$ so the electrons flow from $T$ to
$B$. Note that the highest occupied level of the neutral
nanoparticle is $i=0$ and it is doubly occupied.

In Fig. \ref{Fig_3}a
${\cal C}_B = {\cal C}_T$ so the applied bias
is divided equally between the two contacts. In FB
the first resonance is $0B$ (tunneling from the
$i=0$ neutral nanoparticle level to
lead $B$)  and occurs at $V=50$mV$=2U/e$.
This is followed at higher bias by other
NP's interspersed with CFP's $Q^+$ and $\bar{Q}$. Since ${\cal C}_B =
{\cal C}_T$
the spectrum under RB is similar with the roles of contacts $T$ and
$B$ interchanged;
each NP and CFP under FB has a counterpart {\em at the same voltage} under RB.
The amplitudes of corresponding peaks under FB and RB are not identical because
the calculated $\gamma^{T,B}_i$
are unequal; see Fig. \ref{Fig_2}a.

Experimentally, the $B$ and $T$ tunnel barriers have different
thicknesses and hence unequal capacitances and
very different tunneling efficiencies; see Fig. \ref{Fig_2}b.
Fig. \ref{Fig_3}b shows such a case with
${\cal C}_B=1.25{\cal C}_T$ which implies bias-induced changes in
$\mu^T$ 1.25 times larger than those in $\mu^B$. Thus corresponding NP's
now occur at FB and RB values that differ by the capacitance
ratio. For example, $0T$ is at 45mV in RB while its
FB partner 0$B$ is at 45mV$\times1.25 = $56mV.
Despite the large asymmetry between
$\gamma^{B}_i$ and $\gamma^{T}_i$ in Fig. \ref{Fig_2}b, the amplitudes of
the NP's are quite similar under FB and RB in Fig. \ref{Fig_3}b because
in each case the current must pass through {\em both} the $T$ and $B$ tunnel
barriers. However for asymmetric tunnel barriers the amplitudes of the CFP's
are very different under FB and RB: In Fig. \ref{Fig_3}b the $\bar{Q}$
CFP's are much stronger than the $Q^{+}$ CFP's under FB while the reverse
is true under RB. {\em This is because  asymmetric tunnel barriers affect the
kinetics of positive
and negative charge fluctuations  differently}: A $Q^{+}$ charge
fluctuation implies an electron surplus on the nanoparticle. Under FB this
fluctuation dissipates easily to the drain (the $B$ contact)
since $\gamma^{B}_i$ is large, thus the total current and $dI/dV$ are affected
little by the introduction of an additional decay channel
and the $Q^{+}$ CFP's are weak. Conversely, a $\bar{Q}$
fluctuation (an electron deficit on the nanoparticle) does {\em not}
dissipate as readily
under FB since this requires tunneling from the source ($T$)
contact through the weakly transmitting ($T$) barrier.
Thus here the introduction of a new conducting channel has a larger effect on
$I$ and $dI/dV$ and $\bar{Q}$ CFP's are strong in FB. Under
RB the transport bottleneck is reversed and $\bar{Q}$ CFP's
are weak while $Q^{+}$  CFP's are strong. Thus the new
charge fluctuation resonances that we have introduced here
readily account for the previously puzzling experimental
tunneling features that have no identifiable partner
when the bias is reversed.\cite{ralph_experiments}

In conclusion, we have presented the first calculations of electron
transport through
metal nanoparticles that are based on a microscopic theory of their electronic
structure. We have shown theoretically that charge fluctuations give
rise to tunneling
resonances of a  new type that account for the behavior of previously
unexplained spectral
features that are observed experimentally.

%
%
%
%

This research has been funded by   NSERC and CIAR.

\end{multicols}


\begin{figure}
\caption{Energy levels near $E_F$ (upper panel). Amplitude  of
$\varphi_i$ at the
$T$   and
$B$  surface corresponding to the second level above (right)
and below (left) the Fermi level (center). The second and third
levels above the Fermi energy are nearly degenerate on this
scale.
$\delta^{Al} = (4 E^{Al}_F/3 {\cal  N}^{Al}){\cal V}^{-1}= 5meV$.
$E^{Al}_F$ is the bulk Fermi energy and
${\cal N}^{Al}$ the electron density  of Al.}
\label{Fig_1}
\end{figure}

\begin{figure}
\caption{Tunneling efficiencies for energy levels $|E_F+i\rangle$
$(\bar{i}=-i)$ for
{\em T} and {\em B} leads.}
\label{Fig_2}
\end{figure}

\begin{figure}
\caption{ $dI/dV$ in forward (FB) and reverse (RB) bias on a logarithmic
scale; consecutive ticks indicate a decade. $U=25$ meV, $T_d = 12$ mK;
(a) ${\cal C}_B={\cal C}_T$, (b) ${\cal C}_B = 1.25 {\cal C}_T$. The
inset is a detail of the structure indicated by *.}
\label{Fig_3}
\end{figure}



\begin{references}


\bibitem{reviews_topic}
W. P. Halperin, Rev. Mod. Phys. {\bf 58}, 533 (1986);
J. v. Delft and D. C. Ralph,
Phys. Rep. {\bf 345}, 61 (2001).




\bibitem{ralph_experiments} D. C. Ralph, C. T. Black, and M. Tinkham,
Phys. Rev. Lett. {\bf 74}, 3241 (1995); Physica {\bf 218B},
258 (1996)

\bibitem{nonmag_experiments}
D. Davidovi\'c and M. Tinkham, Phys. Rev. Lett. {\bf 83}, 1644 (1999); Phys.
Rev. B 61, R16359 (2000);
J. R. Petta and D. C. Ralph, Phys. Rev. Lett. {\bf 87}, 266801 (2001)

\bibitem{cobalt_experiments}
S. Gu\'eron {\em et al.}, Phys. Rev. Lett. {\bf 83}, 4148 (1999);
%
M. M. Deshmukh {\em et al.},
Phys. Rev. Lett. {\bf 87}, 226801 (2001)



\bibitem{Agam_PRL_1997} O. Agam, N. S. Wingreen, B. L. Altshuler, D. C.
Ralph, and M. Tinkham, Phys. Rev. Lett. {\bf 78}, 1956 (1997)

\bibitem{canali_PRL_2000} C. M. Canali and A. H. MacDonald, Phys. Rev. Lett.
85, 5623  (2000)

\bibitem{kleff_PRB_2001} S. Kleff, J. v. Delft, M. M. Deshmukh,
and D. C. Ralph, Phys. Rev. B 64, 220401(R) (2001).

\bibitem{aleiner_PhysRep_2002} I. L. Aleiner, P. W. Brouwer and L. I.
Glazman, Phys.
Rep. {\bf 358}, 309 (2002), and references therein.

\bibitem{Bonet_PRB_2002} E. Bonet, M. M. Deshmukh, and D. C. Ralph, Phys.
Rev. B{\bf 65}, 045317 (2002)


\bibitem{NarvaezKirczenow_PRB_2002} G. A. Narvaez and G. Kirczenow, Phys.
Rev. B{\bf 65}, 121403(R) (2002)

\bibitem{mcglynn_book} S. P. McGlynn, L. G. Vanquickenborne, M.
Kinoshita, D. G. Carroll, 
 {\em Introduction to Applied Quantum
Chemistry} (Holt, Rinehart, Winston, New York, 1972)

\bibitem{schiff_book_1955} L. I. Schiff, {\em Quantum Mechanics}
(McGraw-Hill, 1955)


\bibitem{Rippard_PRL_2002} W. H. Rippard, A. C. Perrella,
F. J. Albert, and R. A. Buhrman, Phys. Rev. Lett. {\bf 88}, 046805
(2002)

\bibitem{note_0} Our results are qualitatively similar for
$d^{ox}_B=10\AA$.


\bibitem{russians_metal_grains}
I. O. Kulik and R. I. Shekhter, Zh. Eksp. Teor. Fiz {\bf 68}, 623
(1975) (Sov. Phys.-JETP {\bf 41}, 308 (1975) ); K. K. Likharev, IBM J.
Res. Dev. {\bf 32}, 144 (1988); D. V. Averin and A. N. Korotkov, J. Low
Temp. Phys {\bf 80}, 173 (1990);{\em ibid}, Zh. Eksp. Teor. Fiz {\bf 97},
1661 (1990) (Sov. Phys.-JETP {\bf 70}, 1990);
           D. V. Averin and K. K. Likharev, in
{\em Mesoscopic Phenomena in Solids}, Eds. B. L. Altshuler, P. A. Lee, and
R. A. Webb, Elsevier, Amsterdam (1991).

\bibitem{note_1} We include the bias voltage in this fashion
because by {\em definition} $eV$ is the difference between
the electrochemical potentials in the leads at a given
applied
$V$.

\end{references}
\end{document}